\documentclass[%
 reprint,
 amsmath,amssymb,
 aps,
pra,
]{revtex4-2}
\usepackage{braket}
\usepackage{graphicx}%
\usepackage{dcolumn}%
\usepackage{bm}%
\usepackage{comment}
\usepackage{color}
\usepackage{xcolor}
\usepackage[utf8]{inputenc} %
\usepackage[T1]{fontenc}
\usepackage{tikz}
\usepackage{adjustbox}
\usepackage{url}
\usepackage[hidelinks,colorlinks=true,linkcolor=blue,citecolor=blue]{hyperref}
\usetikzlibrary{quantikz2}

\begin{document}

\preprint{APS/123-QED}
\title{Quantum-Assisted Hilbert-Space Gaussian Process Regression}%

\author{Ahmad Farooq}
\email{ahmad.farooq@aalto.fi}

\author{Cristian A. \surname{Galvis-Florez}}
\email{cristian.galvis@aalto.fi}
\affiliation{Department of Electrical Engineering and Automation, Aalto University, Finland}

\author{Simo Särkkä}
\affiliation{Department of Electrical Engineering and Automation, Aalto University, Finland}

\date{\today}%

\begin{abstract}

Gaussian processes are probabilistic models that are commonly used as functional priors in machine learning. Due to their probabilistic nature, they can be used to capture the prior information on the statistics of noise,  smoothness of the functions, and training data uncertainty. However, their computational complexity quickly becomes intractable as the size of the data set grows. We propose a Hilbert space approximation-based quantum algorithm for Gaussian process regression to overcome this limitation. Our method consists of a combination of classical basis function expansion with quantum computing techniques of quantum principal component analysis, conditional rotations, and Hadamard and Swap tests. The quantum principal component analysis is used to estimate the eigenvalues while the conditional rotations and the Hadamard and Swap tests are employed to evaluate the posterior mean and variance of the Gaussian process. Our method provides polynomial computational complexity reduction over the classical method.
\end{abstract}

\maketitle
\section{\label{sec:level1} Introduction}
Gaussian processes (GPs) are probabilistic machine learning methods widely used in applications such as robotics and control, signal processing, geophysics, climate modeling, financial markets, and data mining, as well as Bayesian optimization and probabilistic numerics \cite{DFR:13:IEEE_TPAMI, SSH:13:IEEE_SPM, HOK:22:Cambridge, DPCSC:15:IEEE_TBE}. GPs are non-parametric probabilistic models that can be used for modeling multidimensional nonlinear functions through their mean and covariance functions \cite{RW:06:MIT,RAS:03:MAC}. However, the traditional GP regression (GPR) methods struggle with computational efficiency, especially when handling large datasets \cite{SS:20:SC}. This limitation becomes particularly pronounced in fields where rapid processing of large-scale data is critical. In this paper, to tackle this challenge, we aim to accelerate the GPR through quantum computing.

The main computational complexity of GPR arises from the computation of the mean and variance of the posterior distribution. This process becomes increasingly computationally heavy with larger datasets, with computational and memory requirements scaling as $O(N^{3})$ and $O(N^{2})$, respectively, for $N$ observations of input data. To alleviate this problem, various methods have been proposed. In the inducing point methods \cite{SZ:05:SGPUPI, RHBSF:21:SGPR, NT:22:NAIPGPR} the covariance matrix is approximated using a smaller number $M$ of inducing points than the full training set, which reduces the computations to $O(NM^{2})$ or $O(M^{3})$ (for likelihood evaluation and prediction, respectively).

In this paper, we concentrate on low-rank methods  \cite{GCRV:10:JMLR, CR:05:JMLR, RR:07:ANIPS} which are based on approximating the precision matrix via a set of $M$ basis functions which also brings the computational complexity down to $O(NM^{2})$ or $O(M^{3})$. In particular, we use the method proposed by Solin and Särkkä \cite{SS:20:SC} which uses the Hilbert space of eigenfunctions defined by a Laplace operator to approximate the covariance function, which offers a tunable balance between computational complexity and approximation accuracy \cite{SS:20:SC, MBASV:23:SC}.

In recent years, quantum computers have emerged as potential replacements for classical computers \cite{PetAl:22:PGOPQT}. They offer exponential reductions in computational complexity for machine learning tasks.  Quantum computing uses the principles of quantum mechanics to implement computational tasks and has been demonstrated for certain types of problems  \cite{BWPRWL:17:Nat}, for example, integer number factoring \cite{S:94:AQCDLF}, fast database search \cite{G:96:FQMADS}, and matrix inversion \cite{HHL:09:PRL}. 

Many quantum algorithms have been proposed for accelerating machine learning tasks. 
Among the plethora of quantum algorithms, the Harrow--Hassidim--Lloyd (HHL) matrix inversion algorithm \cite{HHL:09:PRL} is often used to accelerate machine learning tasks. It serves as the foundation for various other algorithms such as quantum linear regression and quantum support vector machines \cite{RML:14:PRL, WAN:17:PRA, YGW:21:IEEE_TKDE}. 
However, the HHL algorithm has its challenges, for example, in quantum state preparation, unitary simulation, sparsity, and matrix conditioning \cite{Aar:15:NPh}. 

An HHL-based algorithm for quantum-assisted Gaussian-process regression was introduced in \cite{ZFF:19:PRA}. This algorithm assumes that quantum state preparation and unitary simulation can be performed efficiently. Importantly, this algorithm addresses the inherent limitations of the HHL approach by appropriately selecting the covariance function to construct $s$-sparse matrices, and also carefully adjusts the noise parameters to ensure that the matrix remains well conditioned, as indicated by a condition number $\kappa$.  To achieve a desired level of accuracy $\epsilon$, it exhibits a run time that scales as $O(\log(N) \kappa^{2}s^{2}/\epsilon)$. 

The quantum principal component analysis (qPCA) is another quantum machine learning algorithm that draws inspiration from the HHL algorithm to estimate the dominant eigenvalues and eigenvectors \cite{LMR:14:NPh}.
The authors in \cite{CYGYLGL:22:PRA}, proposed a method to prepare the covariance matrix on a quantum computer using annihilation and creation operators and implement the concept of qPCA to approximate the mean and variance of the GPR.
This approach aims to achieve polynomial speedup compared to classical algorithms by overcoming the quantum state preparation and efficient unitary simulation assumptions.

The contribution of this paper is to integrate the Hilbert space approximation of the kernel presented in  \cite{SS:20:SC}, into a quantum Gaussian process regression algorithm.
This approach shifts the prediction complexity from $O(N^3)$ to $O(M^3)$, reducing the dependency from the number of observations $N$ to the number of eigenfunctions $M$ used to approximate the kernel. 

Our methodology begins with the approximation of the kernel function using Hilbert space basis functions on a classical computer. Subsequently, we transfer this data matrix, characterized by a low-rank covariance function, onto a quantum computer. We then apply qPCA for extracting dominant eigenvectors and eigenvalues for non-sparse low-rank matrix into a quantum register. To derive the posterior mean and variance for reduced-rank Gaussian process regression, we employ conditional controlled rotations followed by the Hadamard tests for the mean and the Swap tests for the variance, respectively. We also include numerical examples to demonstrate and validate the effectiveness of our proposed method. Our proposed algorithm shows a polynomial speed advantage over existing classical algorithms for low-rank approximation in GP regression.

The structure of the paper is as follows. In Section \ref{Classical HS-GPR}, we review the classical formulation for the Hilbert space approximation of GPR. We provide the quantum-assisted Hilbert space GPR algorithm in Section \ref{sec:3}. The complexity analysis of the proposed algorithm and its comparison with state-of-the-art methods are given in Section \ref{sec: 4}. Section \ref{sec: 5} discusses the numerical implementation of our algorithm on a classical simulator. We then conclude our findings in Section \ref{sec: 6}.

\section{Hilbert Space Approximation of Gaussian Process Regression} \label{Classical HS-GPR}

In this section, we summarize the classical Hilbert space method for reduced-rank Gaussian process regression (GPR) \cite{SS:20:SC} as well as show how GPR can be rewritten in terms of eigenvalues and eigenvectors. We first briefly review the classical GPR. Then, we show how to approximate the kernel using a Hilbert space of functions defined by the eigenspace of the Laplace operator. Finally, show how to express GPR in terms of singular value decomposition (SVD). This allows us to write these quantities in a suitable form so that they can be calculated using quantum states.

\subsection{Gaussian process regression} \label{sec: GPR}

Gaussian process regression \cite{RW:06:MIT} is a method for modeling and predicting multi-dimensional data. Consider a dataset $\mathcal{D} = {(\mathbf{x}_i, y_i)}_{i=1}^N$, where each $\{\textbf{x}_{i}\}_{i=1}^{N}$ is a $d$-dimensional input vector and $y_i$ is its corresponding measurement. In GPR, we aim to estimate an underlying function $f(\mathbf{x})$ by modeling it as a realization of a Gaussian process. The measurements are then Gaussian distributed with added Gaussian noise $\varepsilon_i \sim \mathcal{N}\left(0,\sigma^2\right)$:
\begin{eqnarray}
   f &\sim& \mathcal{GP}\left(0, k\left(\mathbf{x},\mathbf{x}^\prime\right)\right),\\
   y_{i}&=&f\left(\mathbf{x}_{i}\right)+\varepsilon_i,
\end{eqnarray}
where $k\left(\mathbf{x},\mathbf{x}^\prime\right)$ denotes the covariance function (kernel), which is a positive semidefinite function $k: \Omega \times \Omega \rightarrow \mathbb{R}$. The choice of kernel function drives the quality of the estimation. A common kernel choice for GPR is the square exponential covariance function \cite{RW:06:MIT}:
\begin{equation}
    k\left(\mathbf{x},\mathbf{x}^\prime\right)=\sigma_{f}^{2}\exp\left(-\frac{1}{2\ell^{2}}\|\mathbf{x} - \mathbf{x}^\prime\|^2\right),
\end{equation}
where $\sigma_{f}$ and $l$ are the signal scale and length scale hyperparameters respectively.

The objective in GPR is to predict the mean and variance of the output for new inputs $\mathbf{x}_{*}$. These predictions are derived from the posterior distribution, which is also Gaussian:
\begin{equation}
    p\left(f_* \mid \mathbf{x}_{*}\right) =  \mathcal{N}\left(f_{*}\mid\mathrm{E}[{f}_*], \mathrm{V}\left[f_*\right]\right).
\end{equation}
The mean and variance of the posterior distribution are given by \cite{RW:06:MIT}
\begin{eqnarray}
    	\mathrm{E}[{f}_*] &=& \mathbf{k}_*^{\top}\left(\mathbf{K}+\sigma^2 I\right)^{-1} \mathbf{y},\label{eq: Classic GPR} \\
	\mathrm{V}\left[f_*\right] &=& k\left(\mathbf{x}_{*}, \mathbf{x}_{*}\right)-\mathbf{k}_*^{\top}\left(\mathbf{K}+\sigma^2 I\right)^{-1} \mathbf{k}_* .\label{eq: Classic GPR var}
\end{eqnarray}
Here, we denote by $\mathbf{y}$ the vector with components $y_i$ from the dataset, $\mathbf{K}$ the $N\times N$ matrix with entries $K_{ij} = k(\mathbf{x}_i, \mathbf{x}_j)$ consisting of covariance functions between all input points in the training set, and $\mathbf{k}_*$ is the covariance vector with the $i$th entry being $ k(\mathbf{x}_*, \mathbf{x}_i)$. 
The kernel function can be approximated by a set of basis functions in a suitable Hilbert space as will be discussed next.

\subsection{Kernel function approximation}
We can approximate a kernel function by considering the eigenvalue problem of the Laplace operator \cite{SS:20:SC}:
\begin{equation}
    \left\{
    \begin{aligned}
        -\nabla^2 \phi_j(\mathbf{x})&=\lambda_j \phi_j(\mathbf{x}), && \mathbf{x} \in \Omega, \\ \phi_j(\mathbf{x})&=0, && \mathbf{x} \in \partial \Omega,
        \end{aligned}\right.
\end{equation}
where the domain $\Omega$ behaves well enough so that the eigenfunctions and eigenvalues exist. The functions $\phi_j(\cdot)$ are orthonormal with respect to the inner product 
\begin{equation}
    \int_{\Omega} \phi_i(\mathbf{x}) \phi_j(\mathbf{x}) \mathrm{d} \mathbf{x}=\delta_{i j},
\end{equation}
which also defines a Hilbert space.

All the eigenvalues $\lambda_j$ of the Laplace operator are real and positive.
If the kernel function is isotropic $k(\mathbf{x}, \mathbf{x}') = k(||\mathbf{x}-\mathbf{x}'||)$ then its eigenvalues are given by the scalar function $S(\omega)$, called the spectral density, which is the Fourier transform of $\mathbf{h} \mapsto k(||\mathbf{h}||)$. It turns out that we can approximate the kernel function in the domain $\Omega$ by \cite{SS:20:SC}
\begin{equation}
    k\left(\mathbf{x}, \mathbf{x}^{\prime}\right) \approx \sum_{j = 1}^{M} S\left(\sqrt{\lambda_j}\right) \phi_j(\mathbf{x}) \phi_j\left(\mathbf{x}^{\prime}\right).
    \label{eq: Kernel approximation}
\end{equation}
Using this Hilbert space approximation of the kernel function, we can reformulate the Eqs.~\eqref{eq: Classic GPR} and~\eqref{eq: Classic GPR var}. This modification allows for computationally efficient approximations for the mean and covariance of the GP:
\begin{eqnarray} 
\mathrm{E}\left[f_*\right] & \approx \boldsymbol{\phi}_*^{\top}\left(\boldsymbol{\Phi}^{\top} \boldsymbol{\Phi}+\sigma^2 \boldsymbol{\Lambda}^{-1}\right)^{-1} \boldsymbol{\Phi}^{\top} \mathbf{y} \label{HS-GPR-mean},\\
\mathrm{V}\left[f_*\right] & \approx \sigma^2 \boldsymbol{\phi}_*^{\top}\left(\boldsymbol{\Phi}^{\top} \boldsymbol{\Phi}+\sigma^2 \boldsymbol{\Lambda}^{-1}\right)^{-1} \boldsymbol{\phi}_*,\label{HS-GPR-var}
\end{eqnarray}
where $\boldsymbol{\Lambda}$ is a diagonal matrix with components $\boldsymbol{\Lambda}_{jj} = S(\sqrt{\lambda_j})$, the matrix $\boldsymbol{\Phi}$ has components $\boldsymbol{\Phi}_{ij} = \phi_j(\mathbf{x}_i)$ and $\boldsymbol{\phi}_*$ has components $\phi_j(\mathbf{x}_*)$. We refer to this approximation as Hilbert space approximation for Gaussian process regression (HSGPR) \cite{MBASV:23:SC}.  The approximation of the kernel now depends on the domain $\Omega$ and the set of eigenfunctions chosen in this domain. For the implementation of this paper, we chose $\Omega$ in the domain $[-L, L]$. The Laplace operator in this domain gives rise to the set of sinusoidal eigenfunctions $\phi_j(x) = L^{-1/2}\sin(\pi j(x+L)/2L)$ with their corresponding eigenvalues $\lambda_{j} = (\pi j/2L)^2$. This kernel approximation allows us to reduce the complexity of the matrix inversion needed to find the mean and variance of the GPR.

\subsection{Mean and variance of reduced rank GPR using singular value decomposition}

In this section, we will convert the mean and variance expressions of GPR into a form that enables them to be expressed as expected values of quantum states and calculated in a quantum computer.
Before applying our quantum algorithm, we modify Eqs.~\eqref{HS-GPR-mean} and~\eqref{HS-GPR-var}. For the GPR, we need the eigenvalues and eigenvectors of $\left(\boldsymbol{\Phi}^{\top} \boldsymbol{\Phi}+\sigma^2 \boldsymbol{\Lambda}^{-1}\right)$ which we wish to express in terms of $\boldsymbol{\Phi}^{\top} \boldsymbol{\Phi}$. We need to reformulate in such a way that both quantities share the same set of eigenvectors, allowing us to write the mean and variance of the GPR in terms of this common set of eigenvectors. This will enable us to write these quantities in terms of the expected values of quantum states.

To address this, we define $\mathbf{X} = \boldsymbol{\Phi}\sqrt{\boldsymbol{\Lambda}}\in \mathbb{R}^{N\times M}$, where $\sqrt{\boldsymbol{\Lambda}}$ is a diagonal matrix with elements $\sqrt{\boldsymbol{\Lambda}_{ii}} = \sqrt{S(\sqrt{\lambda_i})}$, which gives
\begin{eqnarray}
         \mathrm{E}\left[f_{*}\right]&=&\mathbf{X}_*^{\top}\left(\mathbf{X}^{\top}\mathbf{X}+
        \sigma^{2} \mathrm{I}\right)^{-1}\mathbf{X}^{\top}  \mathbf{y},\\
         \mathrm{V}\left[f_*\right] &=&\sigma^{2} \mathbf{X}_*^{\top}\left(\mathbf{X}^{\top}\mathbf{X}+\sigma^{2} \mathrm{I} \right)^{-1} \mathbf{X}_*,
\end{eqnarray}
where $\mathbf{X_*^{\top}} = \phi_*^{\top} \sqrt{\boldsymbol{\Lambda}}$. Now the eigenvectors of $\left(\mathbf{X}^{\top}\mathbf{X}+
        \sigma^{2} \mathrm{I}\right)$ are the same as those of $\mathbf{X}^{\top}\mathbf{X}$. 
        
We then begin by applying the SVD to the real data matrix $\mathbf{X}$ which is then expressed as $\mathbf{X}=\mathbf{U}\boldsymbol{\Sigma} \mathbf{V}^{\top}$. Here $\boldsymbol{\Sigma}\in \mathbb{R}^{R\times R}$ is a diagonal matrix containing the real singular values $\lambda_1,\lambda_2, \ldots,\lambda_R$ and the orthogonal matrices $\mathbf{U}\in \mathbb{R}^{N \times R}$ (and $\mathbf{V}\in \mathbb{R}^{R \times M}$) correspond to the left and right singular vectors, respectively.
Taking into account the sum of $\mathbf{X}^{\top}\mathbf{X}$ and $\sigma^{2}\mathrm{I}$, we derive $\mathbf{X}^{\top}\mathbf{X}+\sigma^{2}\mathrm{I}=\mathbf{V}\boldsymbol{\Sigma}^{\prime} \mathbf{V}^{\top}$, where $\boldsymbol{\Sigma}^{\prime}$ is a diagonal matrix with elements $\boldsymbol{\Sigma}^{\prime}_{ii} = \lambda_{i}^{2} + \sigma^2$. Then, the  eigendecomposition of  $\left(\mathbf{X}^{\top}\mathbf{X}+\sigma^{2} \mathrm{I}\right)^{-1}\mathbf{X}^{\top}$ is given by:
\begin{equation} \label{eigendecomposition}
    \left(\mathbf{X}^{\top}\mathbf{X}+\sigma^{2} \mathrm{I}\right)^{-1} \mathbf{X}^{\top} = \mathbf{V}\boldsymbol{\Sigma}^{\prime\prime}\mathbf{U}^{\top},
\end{equation}
where $\boldsymbol{\Sigma}^{\prime\prime}$ has diagonal components $\boldsymbol{\Sigma}^{\prime\prime}_{ii} = \frac{\lambda_{i}}{\lambda_{i}^{2}+\sigma^{2}}$. We can write the Eq.~\eqref{eigendecomposition} as 
\begin{equation}
\label{eigendecomposition2}
    \left(\mathbf{X}^{\top}\mathbf{X}+\sigma^{2} \mathrm{I}\right)^{-1} \mathbf{X}^{\top} =\sum_{r=1}^{R} \frac{\lambda_{r}}{\lambda_{r}^{2}+\sigma^{2}}
    \mathbf{v}_{r} \mathbf{u}_{r}^{\top}.
\end{equation}

Then, the mean of the GPR can be expressed using the SVD as
\begin{eqnarray}
    \label{Mean}
 \mathrm{E}\left[f_{*}\right]&=&\mathbf{X}_*^{\top}\left(\mathbf{X}^{\top}\mathbf{X}+\sigma^{2} \mathrm{I}\right)^{-1} \mathbf{X}^{\top} \mathbf{y}\nonumber\\&=&\sum_{r=1}^{R} \frac{\lambda_{r}}{\lambda_{r}^{2}+\sigma^{2}}\mathbf{X}_*^{\top}\mathbf{v}_{r} \mathbf{u}_{r}^{\top}\mathbf{y}.
\end{eqnarray}

Similarly, we can write the variance of GPR using the SVD as

\begin{eqnarray}
    \label{Variance}
        \mathrm{V}\left[f_*\right]&=&\sigma^{2} \mathbf{X}_*^{\top}\left(\mathbf{X}^{\top}\mathbf{X}+\sigma^{2} \mathrm{I} \right)^{-1} \mathbf{X}_*\nonumber\\&=&\sigma^{2}\sum_{r=1}^{R} \frac{1}{\lambda_{r}^{2}+\sigma^{2}} \mathbf{X}_*^{\top}  \mathbf{v}_{r}\mathbf{v}_{r}^{\top}\mathbf{X}_*.
\end{eqnarray}
We have now expressed the mean and variance of GPR in a form that allows us to compute each of them as expected values of quantum states, which we will do in the next section.

\section{Quantum-assisted Hilbert Space GPR algorithm} \label{sec:3} 

In this section, we propose a low-rank method for quantum-assisted Gaussian process regression which we call quantum-assisted Hilbert-space Gaussian process regression (QA-HSGPR). For its implementation, we have to encode the data matrix $\mathbf{X}^{\top}\mathbf{X}$ into a quantum state. After that, we can implement a quantum algorithm that allows us to extract its eigenvalues. Then, we build the quantum circuits whose expected values correspond to the mean and variance that characterize the GPR.

\subsection{Quantum state preparation from dataset}

Quantum computers encode classical information into quantum states using qubits \cite{NC:11:QCQI}. A quantum state with $n$ qubits can be expressed as a $2^{n}$ dimensional vector $\ket{\psi}=\sum_{i=0}^{2^{n}-1}a_i\ket{i}$, where $\{\ket{i}\}$ represents the computational basis $\{\ket{0\cdots 0}=\ket{0}, \ldots , \ket{1\cdots 1}=\ket{2^n -1}\}$. The coefficients $a_i$ are complex numbers that satisfy the normalization condition $\sum_{i=0}^{2^{n}-1}|a_i|^{2}=1$.  We use the notation $\bra{\psi}$ to represent the conjugate transpose of the quantum state $\ket{\psi}$.

We use an amplitude state encoding scheme to prepare the quantum state \cite{SP:21:SPR}. The amplitude quantum state encoding encodes the classical vector $\left(\alpha_1~\alpha_2~\cdots~\alpha_n\right)^\top$ into the coefficients of the quantum state.
We begin by obtaining a matrix $\mathbf{X}\in \mathbb{R}^{N\times M}$ using the eigenfunction of the Laplace operator in the given domain.
We then vectorize the matrix $\mathbf{X}$ and encode it using amplitude state encoding scheme \cite{SP:21:SPR} as
\begin{equation} \label{Quantum State}
    \ket{\psi_{\mathbf{X}}}=\sum_{m=0}^{M-1}\sum_{n=0}^{N-1}x_{n}^{m}\ket{m}\ket{n}.
\end{equation}
Here, $x_{n}^{m}$ represents the value of the classical data at position $n, m$ of the data matrix $\mathbf{X}$. It is important to note that the entries $x_{n}^{m}$ must satisfy the condition $\sum_{n,m}\left|x_{n}^{m}\right|^2 = 1$.  This ensures that the quantum state is properly normalized.

Encoding data into a specific quantum state $\ket{\psi_{\mathbf{X}}}$ as given in Eq.~\eqref{Quantum State} generally involves a computational complexity of $O(NM)$ in the conventional quantum state preparation methodologies \cite{MVBS:04:PRL,SP:21:SPR}. This complexity measure refers to the total number of quantum gates necessary to achieve the intended outcome. 
An approximate quantum amplitude encoding procedure has recently been proposed for more efficient state preparation \cite{NUSROTTMY:22:PRR}. In that scheme, quantum state preparation is achieved in $O(\mathrm{poly}(\log(NM)))$, when dealing with a real data matrix. The low-rank kernel matrix  $\mathbf{X}$ for HSGPR consists of real-valued entries, which would allow us to prepare it efficiently. %
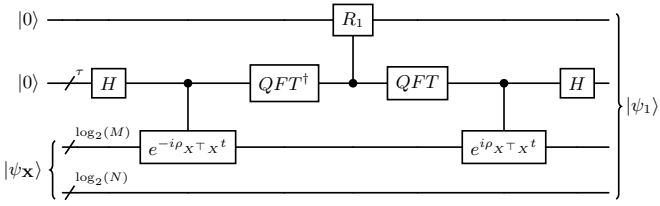
\begin{figure}[tb!]
\centering
\begin{adjustbox}{scale=0.78}
\begin{quantikz}[column sep=0.25cm]
    \lstick{$\ket{0}$}&&&&&&\gate{R_{1}}&&&&\rstick[4]{$\ket{\psi_1}$}\\
    \lstick{$\ket{0}$}&&\qwbundle{\tau}&\gate{H}&\ctrl{1}&\gate{QFT^{\dagger}}&\ctrl{-1}&\gate{QFT}&\ctrl{1}&\gate{H}&\\
    &\setwiretype{n}\lstick[2]{$\ket{\psi_{\mathbf{X}}}$}&\setwiretype{q}\qwbundle{\log_2(M)}&&\gate{e^{- i \rho_{X^{\top}X} t} }&&&&\gate{e^{i \rho_{X^{\top}X}t} }&&\\
    &\setwiretype{n}&\setwiretype{q}\qwbundle{\log_2(N)}&&&&&&&&
\end{quantikz}
\end{adjustbox}
\caption{\label{fig:1} In this figure, qPCA is first employed on the matrix $\rho_{\mathbf{X}^{\top}\mathbf{X}}$  Following this, a conditionally controlled unitary operation is executed based on the eigenvalues register. Finally, we revert the additional $\tau$ qubit register to its original state by executing the corresponding inverse quantum operations to prepare the quantum state $\ket{\psi_1}$.}
\end{figure}

\subsection{Estimation of eigenvalues}
In this section, we show how to extract the eigenvalues of the symmetric matrix $\mathbf{X}^{\top}\mathbf{X}$ and store them in an ancillary quantum register. This allows us to easily perform the conditional rotation operation, which is necessary to obtain the desired amplitude quantities in Eq.~\eqref{Mean} and Eq.~\eqref{Variance}.
 Using the Gram--Schmidt decomposition of Eq.~\eqref{Quantum State}, we can reexpress $\ket{\psi_{\mathbf{X}}}$ as \cite{SSP:16:PRA}
\begin{equation}
    \ket{\psi_{\mathbf{X}}}=\sum_{r=1}^{R}\lambda_{r}\ket{v_{r}}\ket{u_{r}}.
\end{equation}
Let us consider the density matrix $\rho_{\mathbf{X}^{\top}\mathbf{X}}=\operatorname{Tr}_{n}{\ket{\psi_{\mathbf{X}}}\bra{\psi_{\mathbf{X}}}}$ by disregarding the $\ket{n}$ register where $\operatorname{Tr}_{n}$ is the partial trace on the $n$ qubits, which can be written as
\begin{equation}
    \rho_{\mathbf{X}^{\top}\mathbf{X}}=\operatorname{Tr}_{j}\{\ket{\psi_{\mathbf{X}}}\bra{\psi_{\mathbf{X}}}\}=\sum_{r=1}^{R}\lambda_{r}^{2}\ket{v_{r}}\bra{v_{r}}.
\end{equation}
 
Next, we apply the unitary evolution technique of quantum principal component analysis (qPCA) \cite{LMR:14:NPh} $\rho_{\mathbf{X}^{\top}\mathbf{X}}$ to the register $\ket{m}$ of $\ket{\psi_{\mathbf{X}}}$, resulting in
\begin{equation}
 \ket{\xi_1} = \sum_{z=0}^{Z}\ket{z\Delta t}\bra{z \Delta t}\otimes e^{-\dot{\iota} z \rho_{\mathbf{X}^{\top}\mathbf{X}} \Delta t}\ket{\psi_{\mathbf{X}}}\bra{\psi_{\mathbf{X}}} e^{\dot{\iota} z \rho_{\mathbf{X}^{\top}\mathbf{X}} \Delta t},
\end{equation}
for some large $Z$, and the states $\ket{\xi_i}$ are intermediate states along the algorithm. By utilizing the quantum phase estimation algorithm, we can take the $R$ dominant eigenvalues of the operator $\rho_{\mathbf{X}^{\top}\mathbf{X}}$ and write (cf.\ \cite{SSP:16:PRA}) 
\begin{equation}
 \ket{\xi_2} = \sum_{r=1}^{R}\lambda_{r} \ket{v_{r}} \ket{u_{r}}\ket{\lambda_{r}^{2}},
\label{eq: eigenvalues}
  \end{equation}
in which the singular values $\lambda_{r}$ are encoded in the $\tau$ qubits of an extra register.

\subsection{Mean of Gaussian process regression}
In this section, we provide the quantum method for computing the mean of GPR.
We employ the conditional unitary on the ancilla qubit to invert the singular values. We add an extra ancilla qubit. The added ancilla qubit is conditionally rotated based on the eigenvalues register such that
\begin{equation*}
    \begin{split}
      \ket{\xi_3} &= \sum_{r=1}^{R}\lambda_{r}\ket{v_{r}} \ket{u_{r}} \ket{\lambda_{r}^{2}} \left[\sqrt{1-\left(\frac{c_1}{\lambda_{r}^{2}+\sigma^{2}}\right)^2} \ket{0} \right.\\ 
    &\qquad \left. +\frac{c_1}{\lambda_{r}^{2}+\sigma^{2}}\ket{1}\right],
\end{split}
\end{equation*}
where the parameter $c_1$ is chosen such that the quantity $\frac{c_1}{\lambda_{r}^{2}+\sigma^{2}}$ remains upper bounded by 1. After the conditional unitary, we reverse the computation in the $\tau$ qubits register by performing inverse operations of qPCA to bring them back into $\ket{0}$ states
\begin{equation*}
\begin{split}
        \ket{\psi_1} &= \sum_{r=1}^{R}\lambda_{r} \ket{v_{r}}\ket{u_{r}} \ket{0} \left[\sqrt{1-\left(\frac{c_1}{\lambda_{r}^{2}+\sigma^{2}}\right)^2} \ket{0} \right.\\ 
    &\qquad \left.+\frac{c_1}{\lambda_{r}^{2}+\sigma^{2}}\ket{1}\right].
\end{split}
\end{equation*}
The quantum circuit for preparing the quantum state $\ket{\psi_1}$ is shown in Fig.~\ref{fig:1}.  

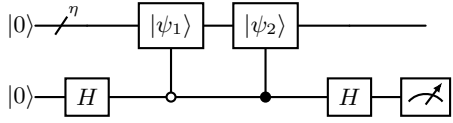
\begin{figure}[tb!]
\centering
\begin{adjustbox}{scale=1}
\begin{quantikz}[row sep={0.65cm,between origins},column sep=0.4cm]
    \ket{0} & \qwbundle{\eta} &\gate{ \ket{\psi_1}}&\gate{ \ket{\psi_2}}&&\\  [0.3cm]
    \ket{0} & \gate{H} &\octrl{-1} &\ctrl{-1}& \gate{H} &\meter{}
\end{quantikz}
\end{adjustbox}
\caption{\label{fig:2} Hadamard test circuit to estimate the mean of GPR.  Here $\eta=\log_{2}(NM)+\tau+1$ qubits. }
\end{figure}

We then prepare another quantum state $\ket{\psi_2}=\ket{\mathbf{X}_*}\ket{\mathbf{y}}\ket{0}\ket{1}$, where $\ket{\mathbf{X}_*} = \sum_{l} {x^l_*} \ket{l}$ and $\ket{\mathbf{y}} = \sum_l y_l \ket{l}$ are normalized quantum states that encode the $\mathbf{X}_*$ and $\mathbf{y}$ vectors respectively. We use the Hadamard test to estimate the inner product between these two states. The circuit diagram of the Hadamard test is shown in Fig.~\ref{fig:2}. 

The implementation of the Hadamard test begins with the application of a Hadamard gate on the ancilla qubit. Depending on the state of the ancillary qubit, different quantum states are generated: $\ket{\psi_{1}}$ for the state $\ket{0}$ and $\ket{\psi_{2}}$ for the state $\ket{1}$.  This results in the composite quantum state
\begin{equation}
\ket{\psi_3}=\frac{\ket{0}\ket{\psi_{1}}+\ket{1}\ket{\psi_{2}}}{\sqrt{2}}.
\end{equation}
Applying the Hadamard gate on the ancilla qubit leads to 
\begin{equation}
      \ket{\psi_3}=\frac{1}{2}\left(\ket{0}\otimes \left(\ket{\psi_1}+\ket{\psi_2}\right)+\ket{1}\otimes \left(\ket{\psi_1}-\ket{\psi_2}\right)\right).
\end{equation}
Both $\ket{\psi_1}$ and $\ket{\psi_2}$ are real vectors, and their inner products $\braket{\psi_1|\psi_2}$ and $\braket{\psi_2|\psi_1}$ are equal. When measuring the ancilla qubit, the probability  $p(0)$ of measuring the ancilla in state $0$ is given by
\begin{equation}
    p(0)=\frac{1}{2}+\frac{1}{2}\braket{\psi_1|\psi_2},
\end{equation}
where
\begin{equation}
\braket{\psi_1|\psi_2}=c_1\sum_{r=1}^{R}\frac{\lambda_{r}}{\lambda_{r}^{2}+\sigma^{2}} \braket{\mathbf{X}_*|v_r} \braket{y|u_{r}}.
\end{equation}
 Thus we obtain an expression equal to the GPR mean as given in Eq.~\eqref{Mean} up to a multiplicative constant. This mean value approximates the output function based on the data points and can be estimated using a quantum circuit.

\subsection{Variance of Gaussian process regression}

In this section, we build the quantum circuit which computes the variance of the Gaussian process regressor.
For that purpose, conditional rotation is applied on the eigenvalues register such that

\begin{equation*}
     \begin{split}
                \ket{\xi_4} &= \sum_{r=1}^{R}\lambda_{r}\ket{v_{r}} \ket{u_{r}} \ket{\lambda_{r}^{2}} \left[\sqrt{1-\left(\frac{c_{2}}{\lambda_{r} \sqrt{\lambda_{r}^{2}+\sigma^{2}}}\right)^2} \ket{0} \right. \\
                &\qquad \left.+\frac{c_{2}}{\lambda_{r} \sqrt{\lambda_{r}^{2}+\sigma^{2}}}\ket{1}\right],
     \end{split}
 \end{equation*}

where the parameter $c_2$ is  chosen such that the quantity $\frac{c_{2}}{\lambda_{r} \sqrt{\lambda_{r}^{2}+\sigma^{2}}}$ remains upper bounded by 1.
We proceed with the algorithm for measuring the ancilla qubit and consider only the measurements in the state $\ket{1}$. Then, discarding the eigenvalue register, ancilla register, and right eigenvector register results in the final state
     \begin{equation}
      \ket{\psi_{1}^{\prime}}=\frac{1}{\sqrt{p(1)}}\sum_{r=1}^{R}\frac{c_2}{\sqrt{\lambda_{r}^{2}+\sigma^{2}}}  \ket{v_{r}},
    \end{equation}
where the probability of acceptance is given by 
\begin{equation}
p(1)=\sum_{r}\left|\frac{c_2 }{\sqrt{\lambda_{r}^{2}+\sigma^{2}}}\right|^{2}.
\end{equation}

\begin{figure}[tb!]

\centering
\begin{adjustbox}{scale=1}
\begin{quantikz}[row sep = {0.8cm, between origins}, column sep=0.25cm]
    \lstick{$\ket{0}$}&&&&&&\gate{R_{2}}&\meter{}\\
    \lstick{$\ket{0}$}&&\qwbundle{\tau}&\gate{H}&\ctrl{1}&\gate{QFT^{\dagger}}&\ctrl{-1}&\\
    \setwiretype{n}&\lstick[2]{$\ket{\psi_{\mathbf{X}}}$}&\setwiretype{q}\qwbundle{\log_2(M)}&&\gate{e^{-i \rho_{X^{\top}X} t}}&\swap{2}&&\\
    \setwiretype{n}&&\setwiretype{q}\qwbundle{\log_2(N)}&&&&&\\
    \lstick{$\ket{X_*}$}&&\qwbundle{\log_2(M)}&&&\targX{}&&\\
    \lstick{$\ket{0}$}&&&&\gate{H}&\ctrl{-1}&\gate{H}&\meter{}
\end{quantikz}
\end{adjustbox}
\caption{\label{fig:var_qc} In this figure, we illustrate the application of qPCA on the matrix $\rho_{\mathbf{X}^{\top}\mathbf{X}}$. Initially, qPCA identifies the eigenvalues and eigenvectors of the matrix. We then apply a conditionally controlled unitary on ancilla register based on the eigenvalue register. Finally, the Swap test is employed to estimate the variance of the GPR. }
\end{figure}
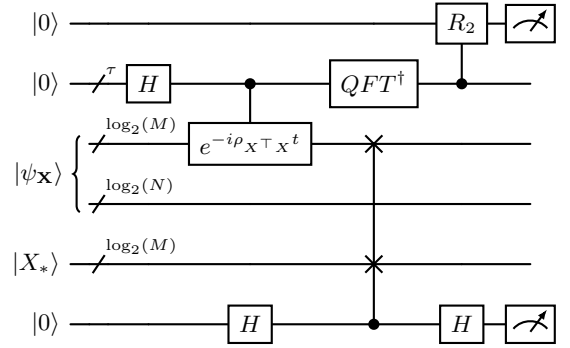

\begin{table}[b]%
\caption{\label{tab:table1}%
The time complexity of each step in the proposed method.
}
\begin{ruledtabular}
\begin{tabular}{ll}
\multicolumn{1}{c}{\textrm{Step}}&
\textrm{Time Complexity}\\
\colrule
Quantum state encoding & $O(\mathrm{poly}(\log(NM)))$ \\
qPCA & $O(\log (M) \epsilon^{-3} )$ \\
Ancilla rotation & $O (\log(\frac{1}{\epsilon}))$\\
Ancilla measurement & $O(\kappa^{2})$
\end{tabular}
\end{ruledtabular}
\end{table}

We use the Swap test to obtain the variance of GPR.  We prepare another quantum state $\ket{\psi_{2}^{\prime}}=\ket{X_*}$.  Using the swap operation between $\ket{\psi_{1}^{\prime}}$ and $\ket{\psi_{2}^{\prime}}$, we can calculate $\left|\braket{\psi_{1}^{\prime}|\psi_{2}^{\prime}}\right|^{2}$ which corresponds to the posterior variance

\begin{equation}
   \left|\braket{\psi_{1}^{\prime}|\psi_{2}^{\prime}}\right|^{2} =  \frac{c_2^{2}}{p(1)}\sum_{r=1}^{R}\frac{1}{\lambda_{r}^{2}+\sigma^{2}}\left| \braket{\mathbf{X}_*|v_{r}}\right|^{2}.
\end{equation}
This is the same expression as we derived in Eq.~(\ref{Variance}), up to a multiplicative constant.
We then multiply with the noise variance $\sigma^2$ to obtain the variance of Gaussian process regressor. Fig.~\ref{fig:var_qc} shows the circuit implementation for computing the variance. %

\begin{table}[b]
\caption{\label{tab:table2}The time complexity of our proposed algorithm against existing quantum and classical counterparts.}
\begin{ruledtabular}
\begin{tabular}{lll}
 Algorithms&\multicolumn{2}{c}{Computational Complexity}\\
&With data &Without data  \\ &loading&loading
\\\hline
 QA-HSGPR&$O(\mathrm{poly}(\log NM) \log M \epsilon^{-3} \kappa^{2} )$ &$O( \log M \epsilon^{-3} \kappa^{2} )$  \\
Zhao \cite{ZFF:19:PRA}&- & $O(\log N \epsilon^{-3} \kappa^{2} )$
\\
Chen \cite{CYGYLGL:22:PRA} & $O(\frac{1}{\sqrt{P_{k}}}dN\log\frac{d}{\delta}\log N \epsilon^{-3}\kappa)$&-\\
\\
HSGPR \cite{SS:20:SC} &\multicolumn{2}{c}{$O(NM^{2})$}
\end{tabular}
\end{ruledtabular}
\end{table}

\section{Complexity Analysis} \label{sec: 4}
In this section, we analyze the computational complexity associated with our proposed method. The algorithm starts with the quantum state preparation step. We employ an approximate quantum encoding scheme to prepare the quantum state $\ket{\psi_{\mathbf{X}}}\in \mathbb{R}^{N\times M}$. This process requires a computational complexity of $O(\mathrm{poly}(\log(NM)))$. Similarly, the preparation of the quantum state $\ket{\psi_2}$ mirrors this complexity.  The total complexity for the preparation of the quantum state is $O(\mathrm{poly}(\log(NM)))$.  

Following the state preparation step, we implement qPCA. The computational complexity for qPCA is $O(\log (M) \epsilon^{-3} )$ where $\epsilon$ denotes the desired error tolerance. The next phase involves a conditional unitary operation, achievable in $O (\log(\frac{1}{\epsilon}))$. However, its complexity is relatively negligible compared with the complexity of qPCA. For both the mean and variance calculations in GPR, the initial algorithmic steps remain same.

To calculate the mean of the GPR, we employ the Hadamard test. The computational complexity of this test is linear in the number of qubits, with measurement accounting only for a constant factor which can be ignore. 
In the variance computation of the QA-HSGPR algorithm, the method involves measurement after the unitary conditional rotation.  This requires $O(\kappa^{4})$ operations on average to measure the ancilla in the excited state. However, applying the techniques of \cite{RML:14:PRL,HHL:09:PRL}, we can reduce this to $O(\kappa^{2})$. Following this, the Swap test is applied which is linear in the number of qubits. The measurement accounts for only a constant factor which can be ignored. Therefore, the overall computational complexity of the GPR  is $O(\mathrm{poly}(\log(NM)) \log (M) \epsilon^{-3} \kappa^{2} )$. A detailed comparison of the computational complexity of each step is summarized in Table~\ref{tab:table1}.   

Classical Hilbert space methods for GPR generally endure a computational load of $O(M^{3})$ \cite{SS:20:SC}.  In contrast, the mean and variance computations of our algorithm demonstrate a polynomially faster speed. 
We also compare our model with that of Zhao \cite{ZFF:19:PRA} whose algorithm complexity is $O( \log (N) \epsilon^{-3} \kappa^{2} )$ dependent on the number of observations $N$ assuming that the data matrix is already prepared in the quantum state. 
If we consider such an assumption, our method would have a complexity $O( \log (M) \epsilon^{-3} \kappa^{2})$, which is primarily dependent on the number of eigenfunctions $M$. 
This shifts the focus in complexity to $M$ rather than $N$ in our model, which significantly reduces the computational complexity, especially in scenarios with large datasets where usually $M\ll N$.

Furthermore, our method demonstrates significant improvements over the recently proposed quantum algorithm for Gaussian process regression. This contemporary model reports a time complexity of  $O(\kappa(\frac{1}{\sqrt{P_{k}}}dN\log(\frac{d}{\delta})\log (N) \epsilon^{-3}+\mathrm{poly}\log (N)))$ where
$P_{k}$ denotes the probability of success for creating the quantum state and $\delta$ indicates the precision of the preparation of the state \cite{CYGYLGL:22:PRA}. This complexity depends on the dimension of the data points, which is not the case for our method. We present a detailed comparison of our method with existing approaches in  Table~\ref{tab:table2}.  This comparison reveals that the overall complexity of our proposed scheme is substantially lower than that of other existing methods.

\section{Numerical Experiments} \label{sec: 5}
In this section, we present the numerical results of our proposed scheme. Our focus is on demonstrating the effectiveness of the method, by performing simulations on a classical computer. We use the built-in Qiskit function for quantum state encoding in our simulation \cite{Qiskit}.

\subsection{Quantum circuit simulation}
Several factors influence the performance of our method. A key aspect is the time parameter in the qPCA algorithm, which we use to estimate the eigenvalues $\lambda_r^2$ in the quantum register, as shown in Eq.~(\ref{eq: eigenvalues}). This estimation is done using the unitary operator $U = e^{- i  \rho_{\mathbf{X}^\top\mathbf{X} }t}$, where we define the time parameter as $t = 2\pi/\delta_R$. Following the phase estimation bounds detailed in \cite{Cleve:1998:QAR},  we can assert that $ \delta_R > \lambda_{\text{max}}^2$, where $\lambda_{\text{max}}^2$ represents the largest eigenvalue of operator $\rho_{\mathbf{X}^\top\mathbf{X}}$. But, for a good approximation of eigenvalues, $\delta_R$ should be slightly greater than $\lambda_{\text{max}}^2$. It could also happen that the qPCA estimation algorithm gives two different approximations to the same eigenvalue. To avoid this problem, we checked the similarity between the different excited states after the qPCA algorithm and discarded the states that likely represent the same eigenvalue in our simulation. The distinguishing of the eigenvalues becomes better when we increase the $\tau$ qubit register.

We then select the dominant $R$ eigenvalues. The selection of the dominant eigenvalues $R$ is a critical factor here. In our demonstration, the selection is made such that the lowest of the $R$th eigenvalues exceeds $0.01$. Specifically, the probability $p$ of finding the desired state, as outlined in \cite{SSP:16:PRA} is bounded by 
\begin{equation*}
    p \leq R\left| \frac{\lambda_{\text{min}}}{\lambda_{\text{max}}} \right|^2.
\end{equation*}
It is important to note that a significant decrease in the smallest eigenvalue will proportionally decrease the probability of measuring the desired state, needing a higher number of shots for an accurate estimation.
We define the constants $c_1$ as $\lambda_r^2+\sigma^2$ and $c_2$ as $\lambda_r\sqrt{\lambda_r^2+\sigma^2}$ in our experiments. 

To precisely mirror the classical results using a quantum computer, a substantial number of qubits and a high number of shots are required. Furthermore, optimizing the hyperparameters is crucial for effective implementation of the algorithm. Our algorithm is ideally suited for fault-tolerant quantum computers. 
\subsection{Simulation results}
To demonstrate the functionality of our method, we have successfully implemented it on a much smaller scale. Our simulation involved $N=16$ data samples derived from an oscillating function in a symmetric length interval $L = 2\pi$, with additive white Gaussian noise $\sigma=0.1$, length scale $l=1$, and signal variance $\sigma_f=1.5$.  We use the $10^{6}$ number of shots for this experiment and $\tau=13$ qubits for the eigenvalue register.
We implement the approximation using a set of sinusoidal eigenfunctions in the domain $\Omega = [-L, L]$ to approximate the kernel. First, we chose $M=4$ and performed estimations for $R = 1, 2,3,4$. We show the behavior of the mean for different $R$. The open-source implementation of our simulation is available in reference \footnote{Source code available at: \href{https://github.com/EEA-sensors/qa-hsgpr-codes}
{\url{https://github.com/EEA-sensors/qa-hsgpr-codes}}}.

\begin{figure}[tb!]
    \centering
    \includegraphics[width = 1\linewidth]{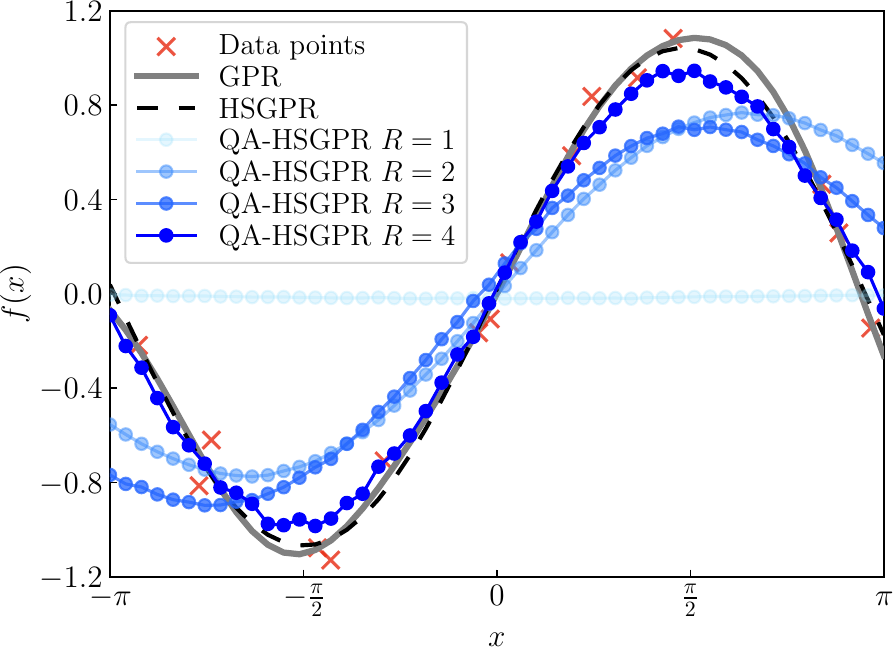}
    \caption{Mean of the GPR using the squared exponential kernel (gray), the Hilbert space approximation of the kernel with $M = 4$ eigenfunctions (black dashed line), and our reduced rank approximation using a quantum circuit (blue lines) with $N=16$ data points (red cross). The blue lines range over $R = 1,2,3,4$ showing how taking a larger rank increases the accuracy of the estimation.}
       \label{fig: mean QHSGPR}
\end{figure}

Fig. \ref{fig: mean QHSGPR} compares the GPR using the exponential kernel, its Hilbert space approximation, and the reduced rank approximation implementing a quantum circuit proposed in this paper. We can see how with $R=4$ the estimation already follows the tendency of the data. However, the estimation result is not exact. There are several reasons for this, first, we are using a limited amount of qubits in the precision of the eigenvalues, which reduces the precision of the mean estimation. Moreover, along the circuit, we have to implement multiple times controlled gates of the unitary operator $e^{\dot{\iota}  \rho_{\mathbf{X}^{\top}\mathbf{X}} t}$ as well as controlled rotations of small angles, which lead to numerical errors in the simulations.

\begin{figure}[tb!]
    \centering
 \includegraphics[width = 1\linewidth]{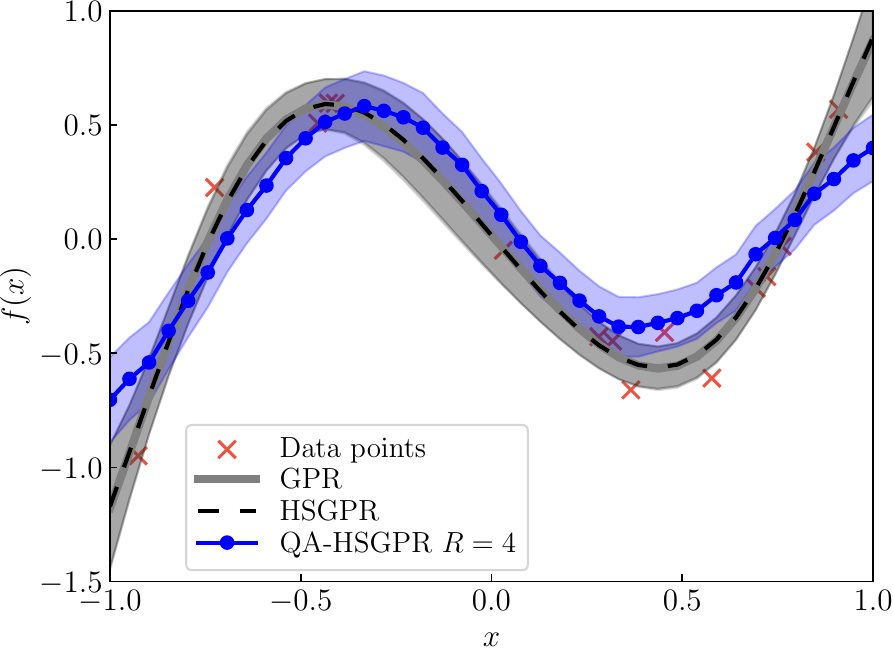}
    \caption{Mean and variance of GPR using the squared exponential kernel (gray solid), the Hilbert space approximation of the kernel with $M = 8$ eigenfunctions (black dashed), and our QA-HSGPR (blue line) with data points $N=16$ (red cross). Each point in the blue line represents a simulation. The shaded areas around each approximation line indicate the $95\%$ confidence intervals, providing a visual representation of the uncertainty associated with each method. We can see that our proposed scheme approximates the HSGPR method well with $R=4$.}
    \label{fig: Quantum HGGPR}
\end{figure}

We also performed another simulation with $M=8$ and the same number of data points with a different function. We performed an estimation with $R=4$, as illustrated in Fig.~\ref{fig: Quantum HGGPR}. The additive white Gaussian noise $\sigma=0.1$, length interval $L=2$, length scale $l=1$, and signal variance $\sigma_f=0.5$.  We use $\tau = 16$ qubits for the eigenvalue register and $10^6$ shots. As can be observed in Fig.~\ref{fig: Quantum HGGPR}, the estimation of the mean and variance of GPR through a quantum computer gives a close approximation of HSGPR. These simulations demonstrate the effectiveness of our algorithm and how it could be implemented when fault-tolerant quantum computers are available. %

\section{Conclusion} \label{sec: 6}

In this paper, we have introduced a novel quantum-assisted Gaussian process regression (GPR) algorithm leveraging a low-rank representation of the GP. Our algorithm addresses the high computational demand of GPR, showcasing how quantum computing can significantly enhance the scalability and efficiency of GPR models. A novel element of our contribution is the incorporation of the Hilbert space basis function approximation \cite{SS:20:SC} into the quantum computing paradigm. This integration leads to significant improvements in computational efficiency, particularly in terms of reducing the computational complexity compared to classical algorithms.  We also provide numerical examples within a quantum setting, which show that the method also works in practice. 

As for future work, probabilistic numerics techniques \cite{HOK:22:Cambridge} provide a means to obtain probabilistic approximations for numerical integrals. A Bayesian quadrature treats the integral as a Gaussian process \cite{HAG:91:JSPI,MIN:00:book}. Being based on Gaussian process regression, Bayesian quadrature is faced with a significant computational challenge when evaluating the integral. The present methodology provides a promising method to evaluate large-scale integrals using Bayesian quadrature on a quantum computer. 

The algorithm proposed here is suitable for fault-tolerant quantum computers, which makes its implementation in NISQ devices a challenge for further work. The complexity of the circuit is mainly dominated by the qPCA and quantum phase estimation algorithm, then, alternative versions of these algorithms can be considered to reduce the complexity of the circuit. For the qPCA algorithm, a hybrid classical-quantum approach that implements a variational circuit can be considered to reduce the depth of the circuit \cite{XCX:21:EQPCAPQC,P:18:QCNISQ}. The previous proposal would reduce the depth of the circuit but increase the classical resources needed to execute the algorithm. On the other hand, it has been shown that iterative approaches of the quantum phase estimation algorithm reduce the complexity of the circuit needed for this task \cite{SBCA:22:IQPE, CCGF:20:OQPE}. The implementation of iterative versions of the quantum phase estimation algorithm would reduce the complexity of the circuit needed to implement our method, enabling the possibility of implementing it in quantum hardware.

\section{Acknowledgements}
We, Ahmad Farooq, Cristian A. Galvis-Florez, and Simo Särkkä want to gratefully acknowledge funding from the Research Council of Finland project 350221.

Ahmad Farooq and Cristian A. Galvis-Florez contributed equally to this work.

\bibliography{apssamp}
\end{document}